# HyperGraphOS: A Modern Meta-Operating System for the Scientific and Engineering Domains

Antonello Ceravola, Frank Joublin
{name}.{lastname}@honda-ri.de
Honda Research Institute Europe GmbH

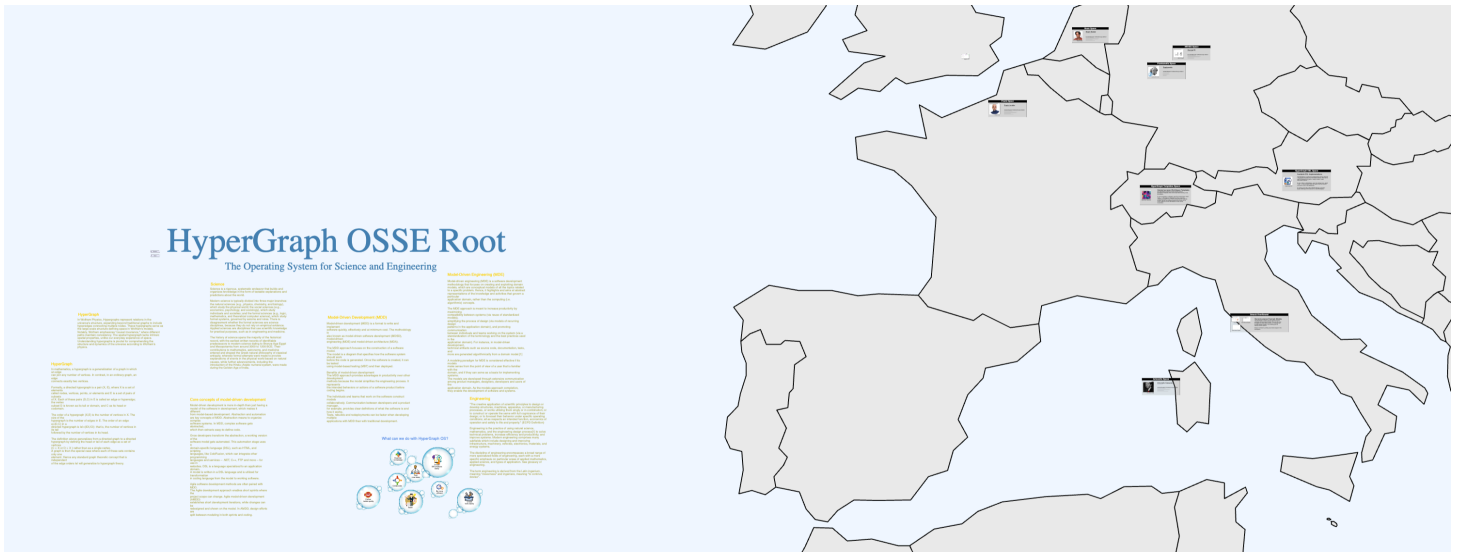

HyperGraph Main WorkSpace

## Abstract

This paper presents HyperGraphOS, a significant innovation in the domain of operating systems, specifically designed to address the needs of scientific and engineering domains. This platform aims to combine model-based engineering, graph modeling, data containers, and documents, along with tools for handling computational elements. HyperGraphOS functions as an Operating System offering to users an infinite workspace for creating and managing complex models represented as graphs with customizable semantics. By leveraging a web-based architecture, it requires only a modern web browser for access, allowing organization of knowledge, documents, and content into models represented in a network of workspaces. Elements of the workspace are defined in terms of domain-specific languages (DSLs). These DSLs are pivotal for navigating workspaces, generating code, triggering AI components, and organizing information and processes. The models' dual nature as both visual drawings and data structures allows dynamic modifications and inspections both interactively as well as programaticaly. We evaluated HyperGraphOS's efficiency and applicability across a large set of diverse domains, including the design and development of a virtual Avatar dialog system, a robotic task planner based on large language models (LLMs), a new meta-model for feature-based code development and many others. Our findings show that HyperGraphOS offers substantial benefits in the interaction with a computer as information system, as platoform for experiments and data analysis, as streamlined engineering processes, demonstrating enhanced flexibility in managing data, computation and documents, showing an innovative approaches to persistent desktop environments. HyperGraphOS bridges the gap between traditional OS functionalities and the technological innovation in UX (User eXperience), together with the evolving needs of modern scientific and engineering applications, thereby providing higher efficiency and productivity compared to traditional operating systems.

HyperGraphOS demonstrates a transformative approach to operating systems by leveraging the unique capabilities of domain-specific languages (DSLs) and a graph-based model representation. This allows users to manage complex data interactions and visualize relational and interconnected information efficiently. Our system's capabilities extend beyond traditional operating systems, incorporating advanced AI integration and modular functionality tailored for complex projects. The platform provides a flexible environment with a user-friendly experience, which, effectively, makes it ideal for a broad set of target domains.

# Keywords

Operating System, User eXperience, Model-Based System Engineering (MBSE), Graph Modeling, Data Containers, Domain-Specific Languages (DSLs), Engineering Applications, Scientific Applications, Web-Based Architecture, Visual Design, Feature-Based Development

# Introduction

Starting with the rudimentary beginnings, the concept of operating systems began to materialize in the 1950s and 1960s with the emergence of the first general-purpose computers. These early systems were manually operated by human system operators who executed programs one at a time **[Bullynck, 2018] [Tanenbaum, 2009][Krzyzanowski, 2014]**. The process involved significant human intervention and lacked any form of automation or multitasking capabilities.

**The Advent of Libraries and Batch Processing**
Initially, libraries of reusable code were introduced to streamline the programming process. This led to the development of batch processing systems **[Korovessi, 2006]**, where a sequence of jobs was processed without human intervention. IBM's 701 and 709 systems are notable examples of early batch processing computers.

**Time-Sharing and Interactive Systems**
The 1960s saw a major leap with the development of time-sharing systems **[Creasy, 1981]**. These systems allowed multiple users to interact with the computer simultaneously by rapidly switching between them. The Compatible Time-Sharing System (CTSS) from MIT and IBM's System/360 were pioneering efforts in this regard.

**Abstractions: Files, Folders, and Shell Commands**
UNIX **[Ritchie, 1974]**, developed in the late 1960s at Bell Labs, popularized pivotal abstractions like files, folders, experimented in Multics and a textual shell with commands. These abstractions provided a simple yet powerful interface for interacting with the system. Files and folders as a means of organizing data was revolutionary. Users could create hierarchical structures to manage their documents efficiently, and use shell commands to execute tasks or automate manipulation of files or navigation in directories. These concepts formed a foundation that operating systems have built upon ever since.

**Personal Computers and Graphical User Interfaces (GUIs)**

The 1980s marked the arrival of personal computers (PCs), bringing computing to individual users. During this era, operating systems like MS-DOS **[Paterson, 1983]** and early versions of Windows **[Carpenter, 2011]** became popular. The introduction of Graphical User Interfaces (GUIs) **[Jansen, 1998]** like Apple's Macintosh **[Williams, 1984]** and Windows revolutionized user interaction. Here, files and folders became visually represented by icons, and the textual shell commands were replaced with graphical elements like desktops, menus, buttons, and windows. However, these visual representations still fundamentally reflect the old abstractions introduced by UNIX to a large public.

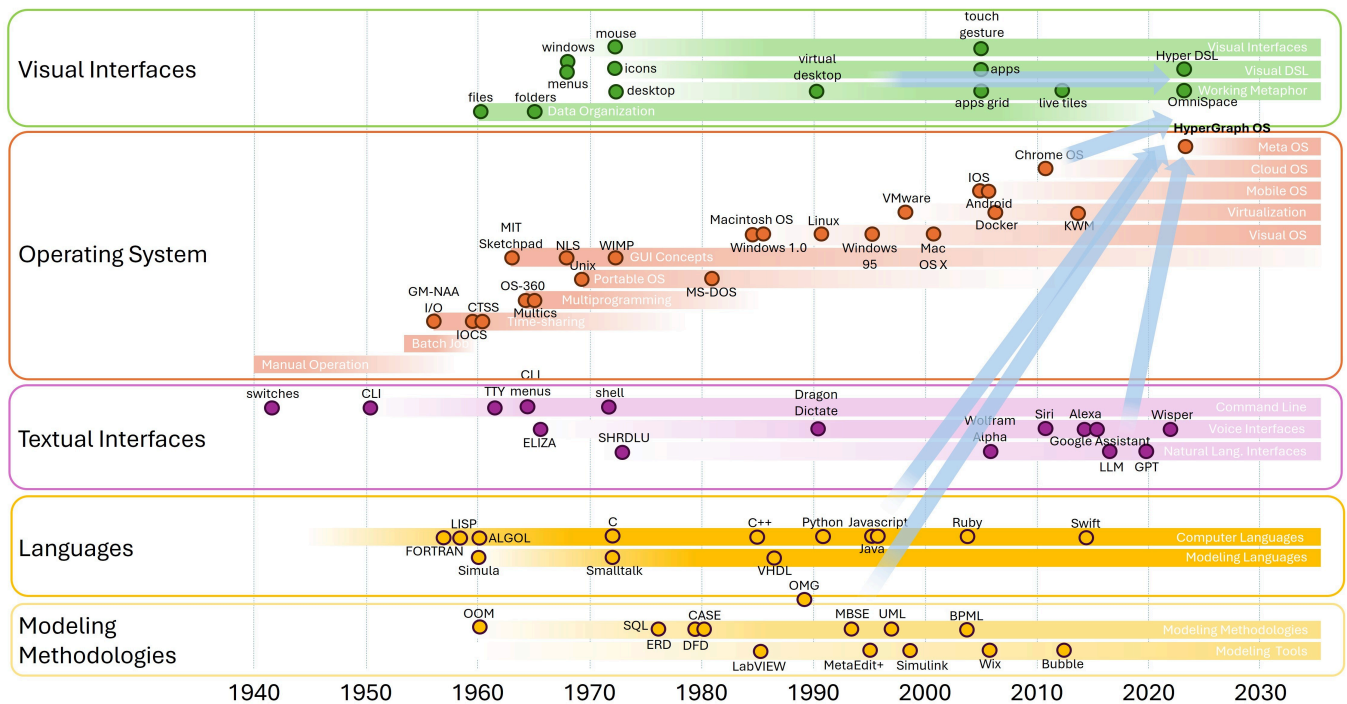

Timeline with major events in the Computer Science domain: Visual Interface related to computer graphical user interfaces; Operating System evolution; Textual Interface that still present and dominating in current operating systems; Languages that had a particular impact in software development; Modeling Methodologies that had major impact in provviding abstractions from computer usage. The blue arrows visualize the main influences that lead to the development of HypergraphOS

**Modern Operating Systems**

In the 2000s and beyond, operating systems have become highly sophisticated, supporting a variety of devices from desktops to smartphones. Despite advancements in technology and user interface design, the core concepts of files, folders, and command-based interactions prevail. Modern systems, including mobile operating systems like iOS and Android **[Wukkadada, 2015]**, or cloud-based and web-based operating systems such as Chrome OS **[Tamayo, 2024]**, has witnessed tremendous advancements over the past few decades, predominantly in response to evolving computational needs and emerging technologies **[Goadrich, 2011]**. Nevertheless, such OS continued to use mostly the same abstractions introduced in the 60' for their user interaction. The focus has shifted to security, multi-core processing **[Wentzlaff, 2010]**, virtualization **[Laadan, 2010]**, and seamless user experiences across devices, but the foundational concepts remain the same.

This chronological account emphasizes the continuity of fundamental abstractions in operating systems from UNIX to present-day systems, illustrating how these concepts have evolved yet remain integral, as created in the early stages, to modern computing.

Despite the overwhelming evolution in computation, data manipulation, user interaction, and the growing range of needs over the years, general-purpose operating systems still fail to account for the specialized functionalities required by several domains in which they are used. For instance, the scientific and engineering fields, which have seen substantial innovations, remain limited by the conventional structure of files and folders. These domains, in particular, demand advanced data manipulation, model design, integration, and interaction with various computational elements and tools. In response to these needs, the development of domain-specific operating systems has gradually gained attention **[Sha, 2022]**.

This paper introduces HyperGraphOS, a novel web-based operating system specifically tailored for scientific and engineering applications. HyperGraphOS find itself at the frontier of Operating Systems, User eXperience, Web Tachnology, Model Based System Engineering **[Wymore, 2018]**, Domain Specific Languages **[Mernik, 2005][Kelly, 2008]**, Data-Flow Architectures **[Culler, 1986]**, Graph Theory **[Bondy, 1976]**, Information Management Systems **[Hicks, 2007]** and Data Science **[Van Der Aalst, 2016]**. By leveraging domain-specific languages (DSLs), a core principle in HyperGraphOS, and graph-based model representations, HyperGraphOS aims to provide a flexible and efficient platform for managing complex models and data in the broadest possible interpretation. This system integrates advanced technologies such as artificial intelligence (AI) components and large language models (LLMs) **[Wei, 2022]**, enabling dynamic interaction with digital information in both interactive and programmatic way.

The motivation behind HyperGraphOS stems from the limitations observed in traditional operating systems when applied to specialized domains as well as the rarity, inadequacy and accidental complexity of existing meta-modeling tools **[Atkinson, 2008], [Ozkaya, 2021]**. Key challenges include the unstructured organization of documents, inefficient data management, and the inability to seamlessly accommodate domain-specific workflows. HyperGraphOS addresses these challenges by offering infinite workspaces, customizable semantic nodes, and visual representation of information, thereby enhancing the user experience and productivity. HyperGraphOS follows a fractal system design **[Lorenz, 2002] [Blair, 2009][Ediz, 2009]**, targeting to be based on a core minumum kernel.

The paper is organized as follows: Section 2 provides a comprehensive background on operating systems and their evolution. Section 3 details the architectural design of HyperGraphOS, followed by an in-depth discussion of its core features in Section 4. Case studies demonstrating the practical application of HyperGraphOS are presented in Section 5. The paper concludes with a comparison to existing systems, highlighting HyperGraphOS's unique contributions and potential areas for future development.

# Background

Operating systems (OS) can be defined from several perspectives, however looking at them from their main function: an operating system is the first program that loads when a computer is powered on, and its primary function is to manage and allocate hardware resources, such as memory, processors, and input/output devices, to ensure that users and applications can interact with the system efficiently **[Tanenbaum, 2009][Silberschatz, 2013]**. As such, OS serves as a program that abstracts the underlying hardware, devices, peripherals, and resources, providing a more user-friendly and efficient interface. Finally, they offer a set of basic, common services that facilitate interactions for both users and other programs. **[Bullynck, 2018]**

There are different types of operating systems which can be broadly categorized into two main groups: general-purpose OS and special-purpose OS **[Hansen, 1973]**. In this paper, we will focus on general-purpose operating systems like Windows, Linux **[Rusling, 1999]**, and MacOS **[Langer, 2003]**, which are designed to be used by anyone, irrespective of their education and knowledge. General purpose operating systems do not require particular technical expertise, allowing user to organize documents and applications without a prescribed structure. These systems are designed to enable users to structure, view, and modify documents through applications specialized on particular file formats.

General-purpose operating systems are employed for a wide range of applications. For example, a household might use a computer to browse the internet, store and organize personal pictures or documents, read news, exchange messages via email or instant messaging, or watch movies and videos. Professionals, on the other hand, may use computers to organize their work, documents, bills, orders, customer data, and processes. Technical users might use computers to create applications or systems with the goal of commercializing their products or making them available to others in various forms. Althought, on one side, general-purpose OS demonstrated their generality and usability in a broad spectrum of domains, on the other side, considering the vast evolution in the technologies and the consolidation of knowledge in each of those domains, general-purpose OS lacks the specificity (based on the consolidated knowledge) one would expects or needs in each specific domain. Looking back at the example above, an household user, with its needs, may be satisfied with the current OS, althogugt may find cumbersome the concept of files and folder, ending-up organizing all documents in the desktop (way more intuitive and accessible than the typical hierarchical file-system structure OSs provides). A professional user, may find the concept of hierarchical folder structure necessary for keeping the work organized, however may have difficulty to relate bills and orders in an intuitive way, and feel forced to use another application (e.g. a spread sheet) to model such relation, something that a file-system is not capable to provide. A technical user may find useful the possiblity to install on a computer source-code editors, compilers and libraries, for creating new application, but find exausting the needs of reorganizing all his windows and documents every time there is the needs to switch to a different project, since OS, currently, have a very poor support for allowing users to organize working environemnts for different projects with different windows and applications, and switching between them on the fly. (for instance, this is a core feature in HyperGraphOS since its first version, and coming now to Windows thanks to a PowerToy extension named App Layout **[Warren, 2024]**).

In this paper we will look at OS from one of the most abstract computer science paradigm: Model Based System Engineering. We argue here that MBSE is the best method we can use to make an analysis and an evaluation of systems, and particularly in this case of Operating Systems. MBSE is focused on models as the primary artifacts in the software development process, which abstract away specific implementation details. To clarify this, lets' break this down in its major constituents:

- *Abstraction*: MBSE emphasizes abstract models of systems, which represent the architecture, behavior, and interactions of software systems without delving into code-level specifics. These models are often formal or semi-formal representations that can be transformed into concrete implementations.
- *Generality*: Because MBSE is centered around models, it can be applied across different domains and development paradigms. Its abstract nature allows it to be layered on top of other methodologies, such as Agile **[Srivastava, 2017]**, Waterfall **[Senarath, 2021]**, or DevOps **[Mohammad, 2017]**, without being bound to the specifics of those approaches.
- *Meta-process*: MBSE incorporates a meta-level approach by allowing the definition of models that can be manipulated, transformed, and analyzed. This meta-process capability supports working with models at different levels of abstraction (from high-level design to low-level implementation), further reinforcing the idea that MBSE is highly general and applicable to various stages of development. MBSE can, in our perspective, be seen as sitting above other methodologies because: it provides a framework for systematically developing systems by using models, which can then be refined into concrete entities or integrated with other processes; it is based on abstractions like meta-modeling, modeling and generations, a process that can even be used in a loop where the result of a modeling process produces the meta-model for the next iteration (multi-level modeling **[Atkinson, 2012]**).

For instance, MBSE could be applied to Agile processes by using models for backlog refinement, sprint planning, or continuous integration pipelines. Moreover MBSE's abstraction level and use of models make it suitable for integration with various other methodologies at different levels. It doesn't conflict with the details of, say, Agile's focus on iterative development, because MBSE focuses on how the system is represented and refined, not on how work is divided or scheduled. Along this line, this paper will examine OS in relation to modeling, specifically model-driven design, model-driven development, and model-based system engineering.

Given the different constituents of an OS, and given the main focus of HyperGraphOS, this paper focus its analysis on how users interfaces and use OSs., It will look at what language an OS provides to users, in terms of domain-specific language (DSL), looking at what users can model with such so-called OS-DSL in relation to the tasks they need or want to perform on a computer. It will also analyse briefly the different innovation proposed in HyperGraphOS. A deeper coverage of each of them will come in a set of follow up papers.

## Operating System DSL (OS-DSL)

By looking at operating systems from a model-based system engineering perspective, ist easy to notice that: an operating system is an application that allow users to model their ideas and needs by positioning elements representing files, folders on the desktop, executing commands through context-menus/taskbar and trough the usage of applications accessible via windows. The OS provides such elements via a domain spcific language (DSL or meta-model). Lets briefly look at those elements. For instance the file element is caracterized by attributes such as name, content, creation/modification dates, access rights, and other OS-specific attributes. Folders elements are specialized files for grouping other files and folders, sharing similar attributes as files. Users place files and folders on the desktop as an easily accessible area on the screen. The desktop is a canvas with a limited space (defined by the screen area of connected monitors), often represented by a folder within the computer's folder structure. Windows elements are used for displaying application's content (usually displayed on top of the desktop). Key attributes for windows elements include size, position, title, state, content and the controlling application. However this OS-DSL is one of the DSL an OS provides. Operating Systems are actually composed of different modules and abstractions, each providing possibility for user to model via a specific DSL. The described DLS has a visual representation (introduced in the transition from text based OS to graphical OS) and a data representation accessible in textual form via applications like compilers and interpreters

OS provide other DSL, for instence, OS gives users the possibiity to execute other programs (than the OS) through a DSL that provides the elements a developer can use to create runnables. Generally, operating system offer such DSL as a set of libraries and APIs. Such API allow developers to create applications, for instance a new text editor, that allow users to manipulate files and folders for particular set of document formats.

OS are also responsible for all peripherals and devices of a computer. Such devices can be integrated in a computer through another DSL that allow OEMs (Original Equipments Manifactures) to interface hardware parts. Along this model-base analysis of OS we notice the plurality of different DSL an OS provides. We will consider the first, being high-level DSL (user interaction), the second mid-level DSL (applications) and the third low-level DSL (hardware).

We do not intend, here, to exaustively cover a full analysis of the different abstractions of an OS (and different DSLs), instead, we will focus here on elements of the high-level DSL introduced above.

In Model-Driven Development (MDD) **[Pastor, 2008]** terms, using an operating system through its GUI is analogous to the modeling phase of the MBSE process. Users create 'models' of their desired content and actions by interacting with UI elements. These models represent the system's intended state or behavior, such as organizing files or configuring settings, which the OS interprets and executes, updating the system state accordingly.

DSLs, as provided by operating systems, play a critical role in enabling users to perform tasks efficiently. They abstract the complexity underlying computer functions, allowing users to focus on their specific activities, leaving details aside. For instance, a developer creating a text editor using low-level APIs does not need to manage hardware resources directly; the OS handles these aspects through its DSL. Similarly, mid-level DSL elements like files and folders enable users to intuitively organize their work without the need to understand file system MTF (Master-File-Table)/i-nodes/FAT (File Allocation Table) concepts **[Puglia, 2019]**. High-level DSLs furnish users with interactive and customizable interfaces, enhancing the overall user experience.

Such DSL is actually grounded in how the real world works, offering a desktop metaphor where files and folders mimic physical documents and folders on a real desk. This analogy helps users intuitively understand how to interact with their digital workspace, representing at best the wanted domain specificity. These analogies bridge the gap between complex computer functions and user-friendly interfaces.

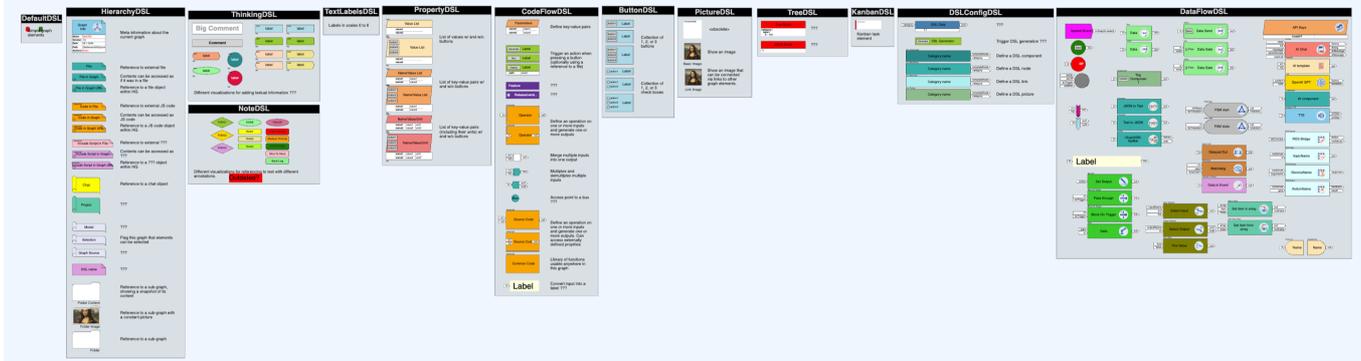

Some of HyperGraphOS available DSLs. We will not be able to cover a description of HyperGraphOS available DSL. However, HyperGraphOS propose a radically different approach to DSLs in relation to the current analysis of OS-DSLs. HyperGraphOS DSL are meant to brake the limitations of file and folders, by providing fine-grained information representation elements, together with the possibliy for user to customize and create their own DSL for modeling their specific application domain.

# Limitation of OS High-Level DSL

Here we review some of the limitations of typical high-level OS-DSL. We do not intend to be exaustive in order to keep paragraphs concised and focused.

1. *Desktop*: common OSs provide a desktop where users can position files and folders and access toolbar and menus for launching applications and commands. One of the main limitations of this DSL element are:

- The desktop area is limited to the computer screen. The number of icons is constrained by the screen size and the layout arrangement the OS offers.
- The visual aspects of icons are limited to small pictures (often determined by the type of files they represent) and their names. The background picture generally has a single image used mostly for decoration purposes (often, with a limited functional role).
- Only the desktop associated with the main screen of a computer is a safe place to be used for positioning icons since additional screens may be optional, making icons on them inaccessible or automatically reorganized by the OS when screens get disconnected. Some OS provide virtual desktops, but they are, generally, limited to the organization of windows (not icons, since they are linked to physical screens).
- Most OS do not recover the state of a computer (or partially recover some of the opened applications and documents) once rebooted, forcing users to reconfigure their computer every time, which leads users to neglect keeping their desktop organized. One solution to that problem was introduced recently in Windows thanks to a PowerToy extension named App Layout **[Warren, 2024]**.
- While programs that extend the capabilities of desktops exist (post-its, drawing, annotations, etc.), their multiplicity and disparity make them unattractive and cumbersome to use for most users.

2. *Files and Folders*: files and folders as a means for organizing documents and applications have several limitations concerning their usability:

- The organization of files and folders depends entirely on the names given by users and the hierarchical structure they choose, generally, a single hierarchy, while multiple hierarchies using symlinks are rarely created. This often leads to inconsistencies and disorder due to varying naming conventions or file placement.
- Adding metadata to files or folders to provide more contextual information is cumbersome and thus rarely used, although it could significantly enhance organization and retrieval.
- Dependencies or relationships between files and folders are typically only possible through the implicit relation of the directory structure, offering limited flexibility.
- Managing large volumes of files and folders can become increasingly difficult without advanced organizational tools.

3. *Windows and Applications*: traditional OS are very application-centric, focusing heavily on the relationship between documents and the applications that visualize or edit them. This structure creates several challenges:

- Data Reusability: the application-centric nature makes it difficult to reuse data across different contexts or applications. For instance, a graphic created in one application may require conversion or export to be used in another, leading to inefficiency.
- Context Switching: users often need to switch between multiple applications to complete a task, complicating the workflow and increasing the cognitive load.
- Consistency: maintaining consistency in document formatting and functionality between different applications can be challenging, as each application may offer unique features and limitations.
- Resource Management: applications often manage their resources independently, which can lead to redundant data storage and inefficient use of system resources.
- Lastly, the heavy reliance on graphical user interfaces (GUIs) can become a significant limitation regarding automation and seamless integration with other advanced tools and systems. Many of the high-level domain-specific languages (DSLs) embedded within traditional operating systems are not inherently designed for easy programmability, thereby limiting the ability to automate repetitive tasks or integrate with external automated processes and systems (MacOS being an exception here, with its Automator **[Apple, 2024]** now replaced by the Workflows and Shortcuts app).

# The HyperGraphOS Concept

HyperGraphOS is built on a fractal architecture that allows the scalable expansion of both capabilities and resources, starting from a small set of core functionalities. This concept defines our MetaOS. In this brief overview, we will outline some of the features available in HyperGraphOS, without going into detailed explanations. Additional information will be provided in future publications.

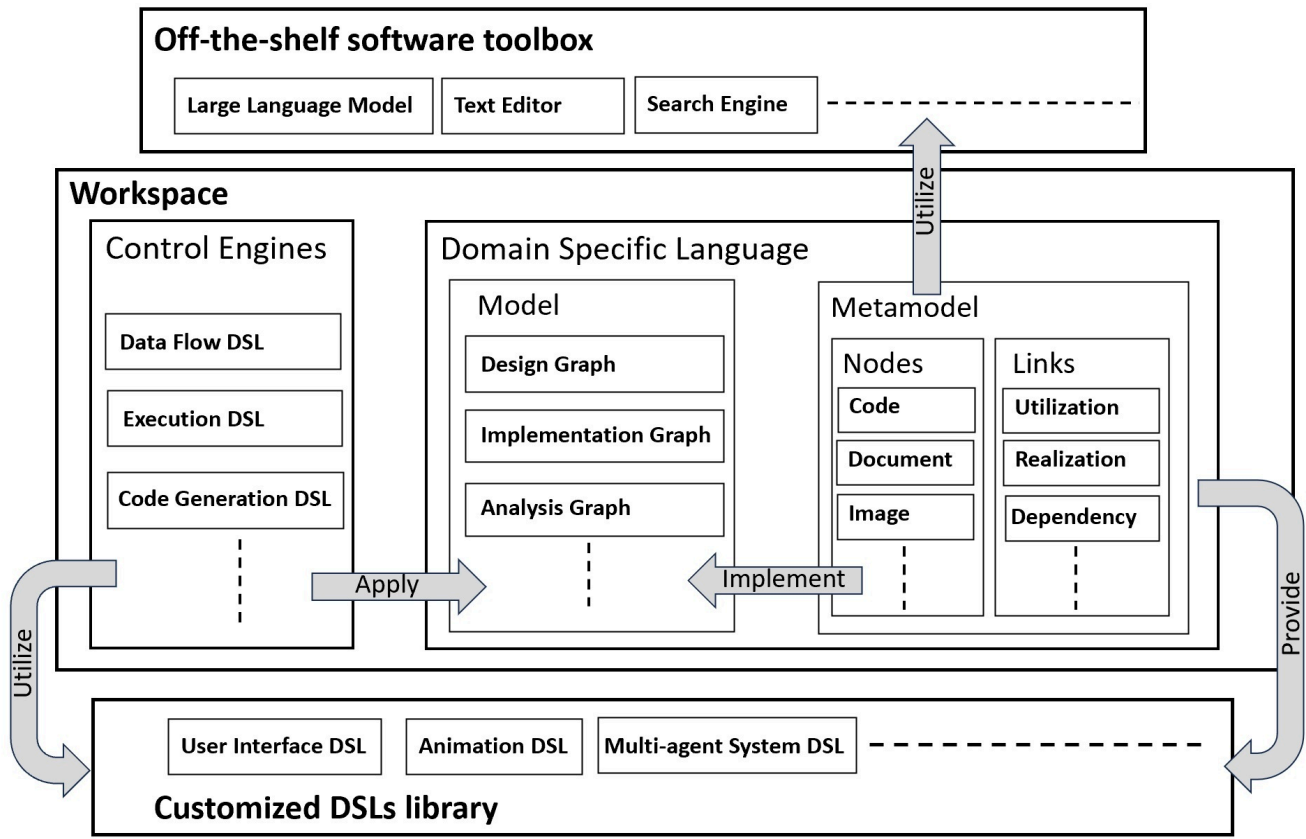

Main HyperGraphOS Concepts. The concepts of WorkSpace with DSLs, meta-models, models and Control Engines are the main core elements of HyperGraphOS UX. It uses several off-the-shelf software toolbox and Customized DSL libraries (as meta-meta-models for DSLs).

HyperGraphOS is designed to transform how we interact with computers and digital information systems. By utilizing DSLs, HyperGraphOS allows files and documents to represent their content and attributes in a way that is tailored to their specific domain. This innovative approach enables the creation of nodes that naturally link and visually represent data, turning traditional file management into an interconnected web of information that is easily customizable, annotatable, and maintainable.

**Unified Documents and Nodes**
HyperGraphOS introduces DSLs to encapsulate both the data and visual aspects of documents. Nodes (representing files) are inherently linked to other objects, similar to the Unix symbolic link concept, allowing easy access to the same file from multiple locations. Nodes in HyperGraphOS can represent a wide range of semantics, including programs, numbers, workspaces, images, texts, and more. Links define specific relationships, such as dependencies, causal connections, and interactions, providing a rich context for data relationships. Each node and link can be associated with specific editors and viewers based on its MIME type, enabling seamless interaction with various types of data.

**OmniSpace: Visual Organization and WorkSpaces**

HyperGraphOS eliminates the need for traditional directories. Instead, documents can be grouped visually by placing them next to each other, using container nodes, linking them with visual connectors, or organizing them within WorkSpaces. WorkSpaces can be thought of as virtual machines, each with its own data and programs, and with the ability to switch between them. WorkSpaces are infinite, flexible environments where documents and applications for specific tasks co-exist in an organized way, with connections to other WorkSpaces (local or remote) in what is called OmniSpace. OmniSpace contrasts with traditional desktops and file systems, offering an expansive and dynamic working environment with limitless possibilities. The system saves the state of all WorkSpaces and navigation history, ensuring continuity in work processes and enabling users to resume tasks exactly where they left off.

**Distributed and Flexible Architecture**

WorkSpaces and node content can be distributed across various storage solutions, including local storage, cloud services, remote computers, and devices. This flexibility enables diverse representations and applications of WorkSpaces. Users can create custom DSLs to represent documents, links, and groups in the most intuitive way, supported by a Meta-DSL for creating tailored languages. HyperGraphOS also provides tools for generating documents from existing ones, supporting model-based systems engineering and development.

**Advanced Command Shell and AI Integration**

JavaScript **[Flanagan, 2007]** serves as the default shell language in HyperGraphOS, enabling the programmatic manipulation of nodes and links. The "Developer Tools" environment in browsers facilitates this interaction. HyperGraphOS includes a powerful find function that goes beyond traditional search capabilities, allowing for complex search expressions with mathematical operators, as well as a smart search feature that integrates web and AI search, making information retrieval efficient and effective within the current workspace. A global search across workspaces is planned for a future release. Additionally, an integrated AI assistant within each document provides on-demand assistance and maintains a persistent reference to AI interactions, enhancing user content creation and maintenance.

**Minimalistic and Efficient Design**

Despite its robust capabilities, HyperGraphOS maintains a small system footprint with a minimalistic user interface (e.g., a single contextual popup menu), ensuring ease of use and intuitive interaction. The concept of applications is reimagined as open and modular constructs, moving away from the monolithic executable model. This allows for a more flexible and integrated approach to application development and use. HyperGraphOS represents a paradigm shift in digital information management, offering a powerful, flexible, and intuitive system for users to interact with data in more meaningful ways.

Nodes may represent complex data structures (e.g., code, JSON data, images, videos, etc.). HyperGraphOS provides a set of viewers and editors (which users can extend) to visualize or edit the content of such data structures (e.g., source code editors for multiple languages, web viewers, rich text editors, etc.).

**Pre-Defined DSL**

We have developed a series of DSLs in HyperGraphOS, some of which have been used in actual projects we've worked on over the past years, while others were created for demonstration purposes. The Dialog DSL was developed to model the behavioral system of a virtual avatar receptionist (for more details, see the section: Case Study). The UserInterface DSL was created to quickly build user interfaces for controlling the setup and execution of various systems. The CodeGeneration DSL is one of the core DSLs of HyperGraphOS, enabling low-code code generation for users. The CodeFlow DSL was developed to challenge traditional programming paradigms and has been used to reprogram HyperGraphOS with itself. The Dataflow DSL is currently the most widely used for rapid prototyping of scientific projects, particularly in the field of AI. Various other DSLs have been created for brainstorming, project management, and more.

**User DSL**
HyperGraphOS has been designed with the goal of empowering users to better express their needs. To support this, we have integrated the ability to extend the available DSLs. Users can define their own DSL to achieve the most effective representation for their specific problems or tasks. To minimize the learning curve, we have developed a DSL specifically for creating new DSLs. This minimalist approach provides a simple and intuitive method for crafting custom DSLs. Users can leverage a set of configurable meta-models to design visual DSL elements. Recognizing that the need for DSL creation can vary greatly, we also offer the option to define custom meta-models, allowing users to build any visual DSL they envision.

**iMBSE - Intelligent Model Based Software Engineering**
In HyperGraphOS, we are introducing a novel modeling process supported by AI. The concept is based on an AI assistant skilled in MBSE, designed to assist users throughout the modeling process. The AI assistant has access to the current model the user is working on and can read from and write to it by adding or removing nodes and links. It can also suggest improvements and propose implementation details for the code generation phase. The AI assistant is intended to serve as an advisory partner to the human user, facilitating and supporting the modeling process in a collaborative manner.

**Multi-Level Modeling**
In HyperGraphOS, we provide an infinite workspace that allows the creation of large-scale models composed of a family of models or sub-models. Within such a model, the same components can simultaneously serve as a Meta-Model, Model, Application, and Deployment Model. Although the concept of multi-level modeling is still in its early stages [Atkinson, 2014], we have observed that HyperGraphOS facilitates a natural approach to designing and implementing models that span different levels of abstraction within the same model graph. This represents an area for future investigation and experimentation.

**Models@Run-Time**
The Dataflow DSL **[Agerwala, 1982]** was developed concurrently with an execution engine designed for it. The execution engine can be used during the modeling process to make data flow models executable. Together, the Dataflow DSL and its engine form a model for data flow and its execution. The engine operates alongside the model, adhering to the data-flow paradigm. In addition to supporting model execution, the engine can assist in the code generation process for the data flow model itself. Several examples within HyperGraphOS are provided to illustrate its applicability and flexibility. With this engine, we aim to inspire users to create additional engines that can be applied in different contexts and for various purposes.

**The Animator**
HyperGraphOS provides a very light animator feature, so flexible that it can be used in a large set of contexts. One of the usage we have explored is in combination with the DataFlow Engine (DFE). During the execution of a model through the DFE, the engine can keep track of the flow of execution. Such flow can be used in the animator to replay back the execution order of the model. The animator data format allows to add annotations, insert pause or change the execution speed at any time. Given this flexibility, for instance, it is also possible to write own animation sequence that automatically navigate through WorkSpaces, and by using the integrated Text To Speech (TTS), read alowd a predefined text. We used this feature to create guided tours for illustrating the content in HyperGraphOS.

**Feature-Based Code Generation & Testing**
As briefly mentioned above, we introduced a new paradigm for software development in HyperGraphOS,

consisting of the following elements: a CodeFlow DSL, a Deployment DSL, a template-based code generation process, and a CodeFlow modeling design principle. These elements enable the design and development of software systems with three main capabilities: first, they allow the initial simple idea of software components to be maintained throughout their evolution within the full system; second, they facilitate the definition of a feature graph that supports product-line development; and third, they enable the design, development, and testing of the entire application system by organizing the model across two dimensions: vertically, by organizing the system into features, and horizontally, by modeling each feature in a testing sub-model. This sub-model allows for the easy debugging and modification of features in isolation, independent of the full system, while keeping feature integration active across the entire system. This unique modeling paradigm will be the subject of a future publication.

**Model-Based Deployment System**

HyperGraphOS natively supports the design, development, and deployment of systems. We propose a model-based deployment system that models both the source system and the destination deployment location. This system is based on the CodeFlow concept, extended with the necessary deployment functionality, such as the transfer of target objects, executables, and data. Additionally, it provides the capability to remotely control the execution of the deployed system on the server side.

**Collaboration**

We propose the integration of multi-disciplinary collaboration tools within HyperGraphOS to support collaboration in the modeling process among people with diverse backgrounds. While these tools are still in their early stages at the time of writing this paper, they are designed to enable seamless communication and coordination among team members, facilitating the exchange of ideas and the division of labor based on individual expertise. By incorporating collaborative features such as real-time editing, version control, and interactive annotations, HyperGraphOS aims to create a synergistic environment where diverse perspectives can converge, ultimately enhancing the overall quality and efficiency of the modeling process.

**HyperNetworks**

HyperGraphOS is deeply integrated with web technology, with the concept of many of its elements (such as DSLs) based on Universal Resource Locators (URLs). As a result, networking across different computers is a fundamental aspect of HyperGraphOS. For example, WorkSpaces are accessed via URLs, making it natural to navigate between WorkSpaces, which may be distributed across an accessible computer network. Similar to navigating web pages through hyperlinks, HyperGraphOS offers a comparable navigation concept among WorkSpaces. This feature presents opportunities that have yet to be fully explored.

# System Architecture

HyperGraphOS is composed of three main modules:

- *Kernel*: This is the main hardware abstraction layer that HyperGraphOS runs on. It can be composed of a Linux kernel or any other operating system kernel capable of supporting the CPU, memory, disk, and input/output devices. The kernel also provides a web-browser sub-service used to run the front-end of HyperGraphOS.
- *Back-end*: This module serves as the interface between the kernel and the front-end. Built on NodeJS **[Basumatary, 2022]**, it is responsible for providing the data representing workspaces/files and executing server-side scripts.

- *Front-end*: This represents the user interface of HyperGraphOS. It runs within the web-browser sub-service provided by the kernel. The front-end features a single, full-screen graphical canvas that displays graphs managed through the commercial library GoJS. Each graph represents a workspace, equivalent to a computer desktop (or a virtual machine), and is stored in the kernel file system as a JSON file **[Smith, 2015]**.

The system employs JavaScript for both the client and server sides, utilizing several key open-source libraries. These include Ace.js for handling source code and text editors, Explore-editor.js for rich text documents, and Winbox.js for managing windows and dialogs, among others.

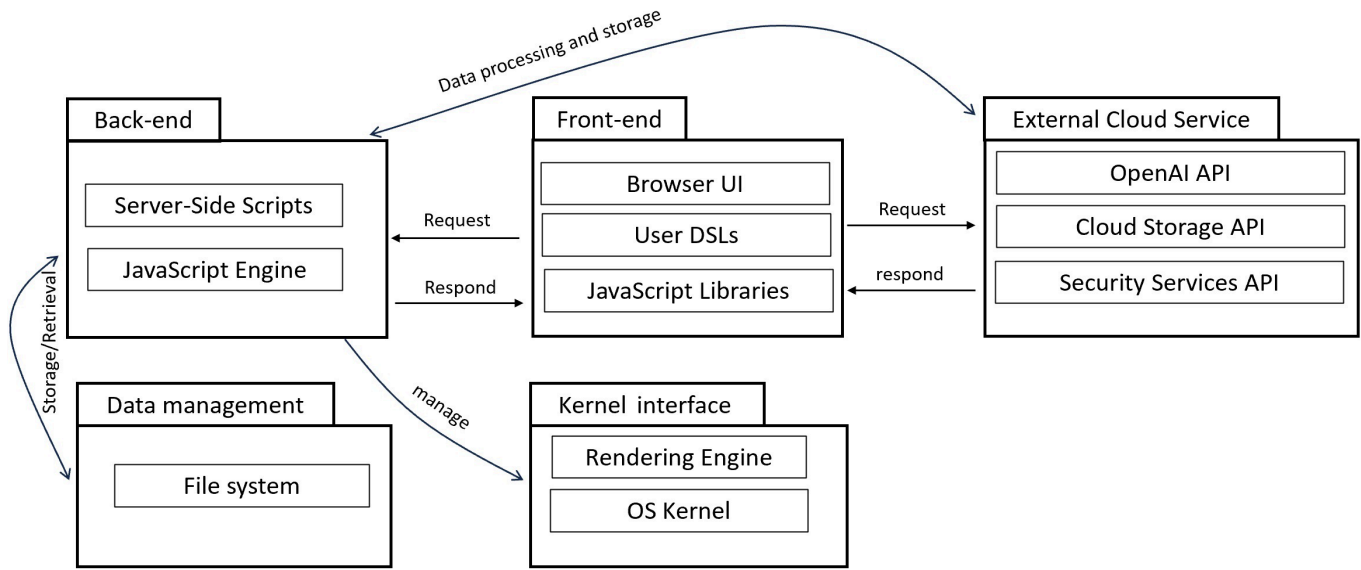

Main HyperGraphOS Architecture Diagram. The starting point for UX is the Front-end, which is based on a web client running in a browser and providing full interaction with users. It utilizes the Back-end as an intermediary to access the file system or execute scripts on the server. The Front-end may also access external cloud services for specific tasks (e.g., AI requests to OpenAI models) or for navigating the HyperNetwork (see previous sections). HyperGraphOS is built on a basic Kernel that manages all machine access and device operations.

Instead of using traditional databases, HyperGraphOS relies on a custom organization of files and directories on the server side for data storage. This approach simplifies the storage solution but may require future enhancements to address more complex data handling needs.

Scalability is inherently managed by design, utilizing the flexible and distributed nature of JSON files that represent workspaces. This approach ensures that multiple nodes or workspaces can be handled efficiently without the need for extensive additional infrastructure.

Security and privacy are managed through external services, ensuring that user data is protected while keeping the primary system architecture lightweight and unobstructed.

For integration and APIs, HyperGraphOS provides APIs that allow programmatic navigation and modification of graphs, with each graph represented as a JSON object. Both the front-end and back-end offer libraries to assist users in defining code generators for their applications, ensuring smooth integration with other tools and systems.

# Case Studies

# Case Study 1: Virtual Receptionist for Visitor Registration

This case study focuses on the development of a virtual receptionist system used for visitor registration at our research institute.

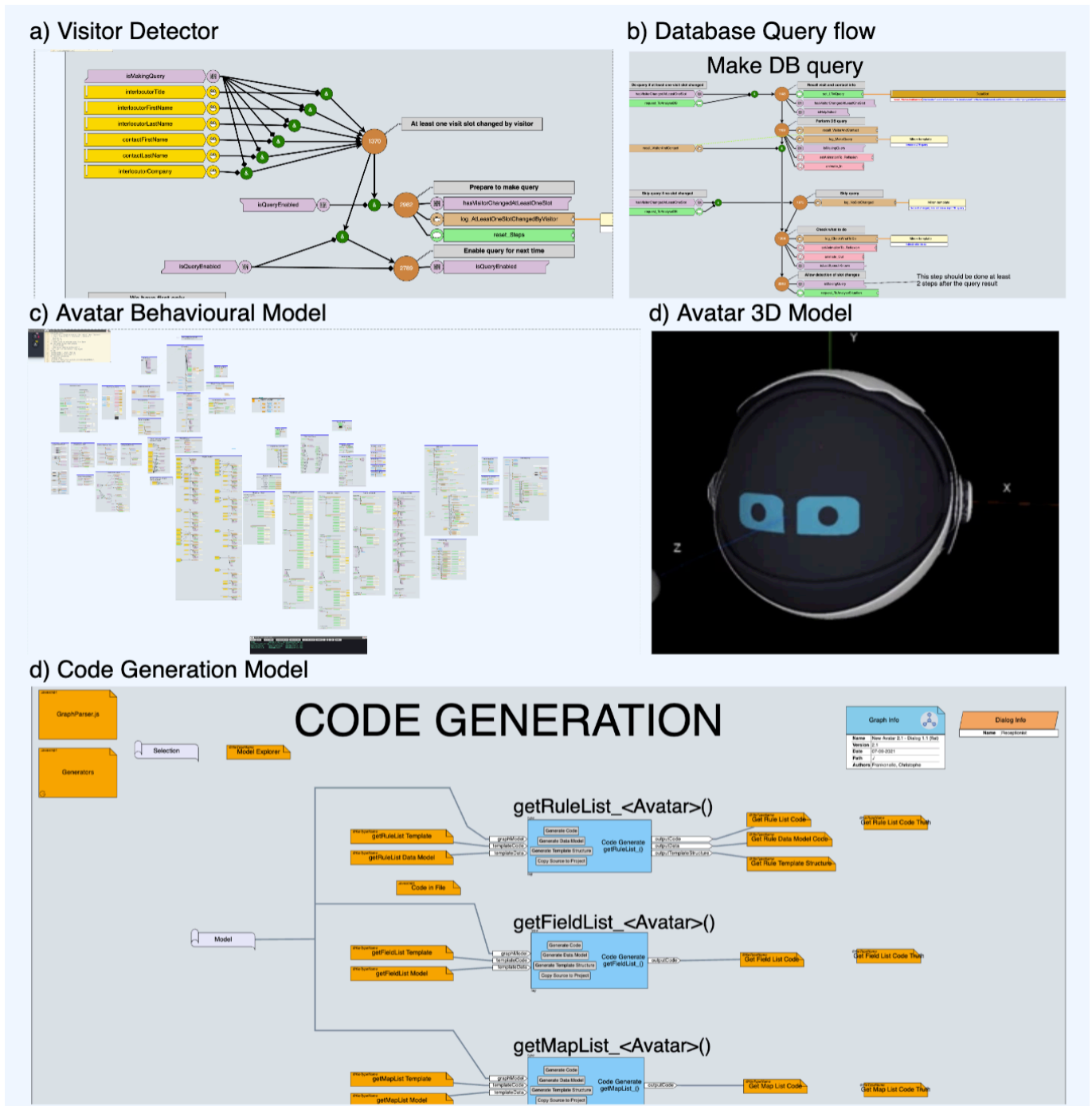

Different view on the Avatar Receptionist Model

This system was initially developed before the widespread adoption of large language models and was originally built using a recursive neural network to define a behavior engine for the AI of the receptionist. The paper "Introducing Brain-like Concepts to Embodied Hand-crafted Dialog Management System" **[Joublin, 2024]** discusses the challenge of creating effective dialog systems capable of seamlessly interacting with humans through natural language or speech. Traditional dialog systems often face issues such as the need for large amounts of training data, difficulty in defining appropriate reward functions, and limited controllability and explainability.

To tackle these issues, the authors designed a neural behavior engine inspired by neurobiology and neuropsychology, incorporating concepts such as mirror neurons and embodiment. This system allows for the creation of mixed-initiative dialog and action generation based on hand-crafted models using a graphical language. The implementation was demonstrated through a virtual receptionist application running in a semi-public space, showcasing its ability to handle real-world interactions with users.

Specifically, HyperGraphOS played a crucial role in two key aspects of the development:

- HyperGraphOS was utilized to define a domain-specific language (DSL) for implementing the behavior engine. In a dedicated workspace, a clock-based architecture model was designed, and code was generated into a target library used by the avatar receptionist system.
- HyperGraphOS was employed to design a DSL representing dialog based on a parallel state flow. This model comprised 4246 nodes and 3890 links, which were used to generate various data files, such as dictionaries (4033 generated lines), weights for the recurrent network (4659 generated lines), and Natural Language Processing (NLP) **[Chowdhary, 2020]** intents (2410 generated lines). Average code generation time (from reading the model to generating all files) is less than 3 seconds.

The application of this brain-like inspired architecture showed several benefits. The virtual receptionist was able to manage dialog states and context effectively, leading to improved user interactions and satisfaction.

The system demonstrated robustness in real-time scenarios, efficiently switching between different modalities (e.g., speech, text, facial expressions) and handling errors gracefully. The use of a domain-specific language (DSL) for behavior modeling facilitated the development process, ensuring scalability and ease of maintenance. The user study confirmed high usability and acceptance, highlighting the system's potential in enhancing human-machine interactions through more natural and intuitive dialogs.

## Case Study 2: Robotic Task Planning and Execution

This case study describes a robotic project where HyperGraphOS was employed to design a robotic control system for task planning and execution using a dataflow DSL **[Joublin, 2023]**. Our institute has a long history of using visual DSLs and data-flow frameworks in robotic research **[Ceravola, 2006]**, **[Sadik, 2019]**. The model was created in a dedicated workspace and implemented a multi-agent system based on OpenAI's large language models. The system took user requests in natural language and transformed them into a plan of robot actions. These actions were first simulated through a robotic simulator and then executed by a real robot in the physical world. The scenarios the robot could handle included preparing pizza, making cocktails, and stacking cubes. The system was built using a model with 414 nodes and 355 links. With HyperGraphOS, a functional version of the system was realized within two weeks. Furthermore, HyperGraphOS's execution engine enabled the running and debugging of the entire system within the workspace, significantly streamlining the development process.

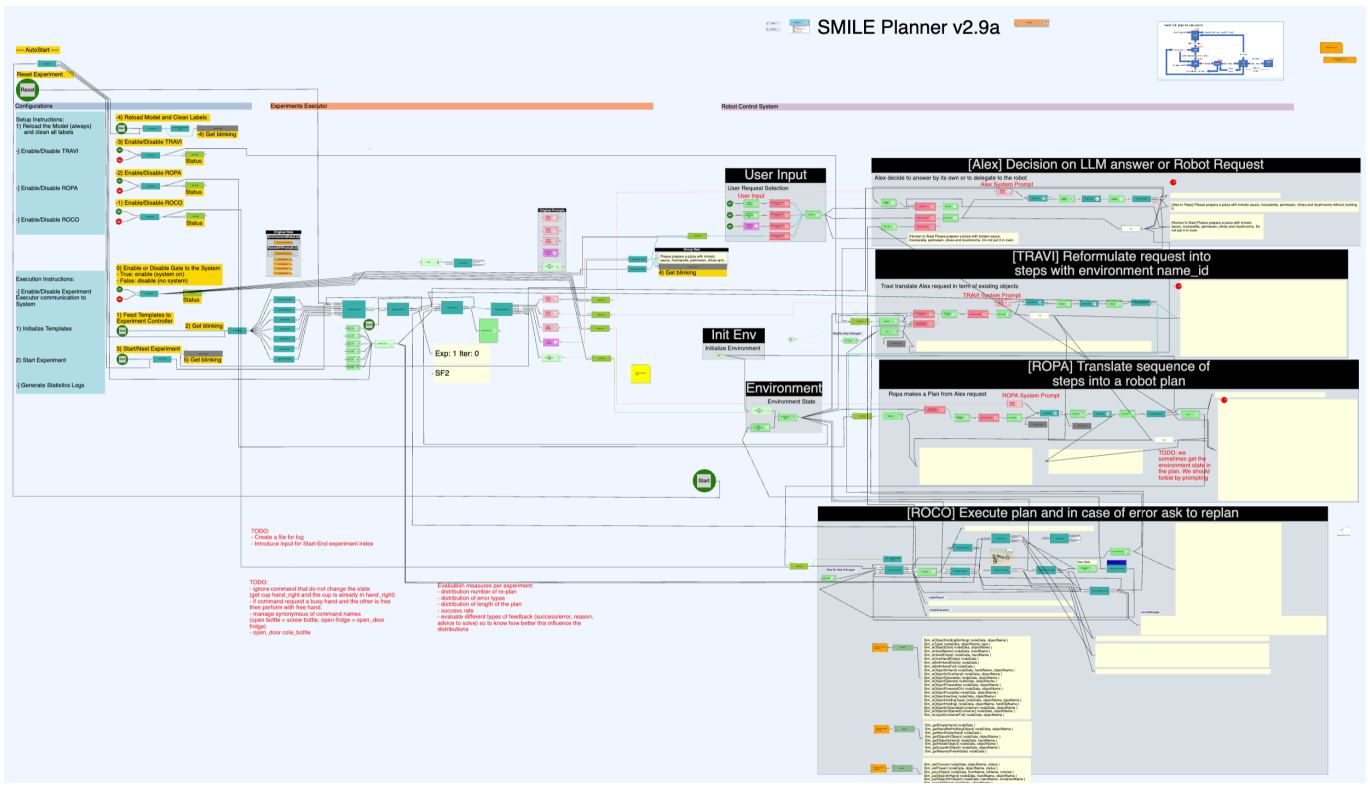

Full Model of the CoPAL Multi-Agent System

The CoPAL system addresses the complexity of enabling fully autonomous robots to operate and adapt in dynamic, open-world environments. Traditional task and motion planning methods often struggle with handling logical, semantic, and physical errors, and they lack the ability to effectively incorporate natural language instructions and low-level feedback into the planning process. CoPAL integrates Large Language Models (LLMs) into a hierarchical architecture that manages multiple cognitive levels—reasoning, planning, and motion generation. It features a novel closed-loop task planning mechanism with multi-level feedback loops, enabling the system to adapt and correct plans in real-time. This architecture was tested in both simulations and real-world scenarios, including blocks world, barman, and pizza preparation tasks.

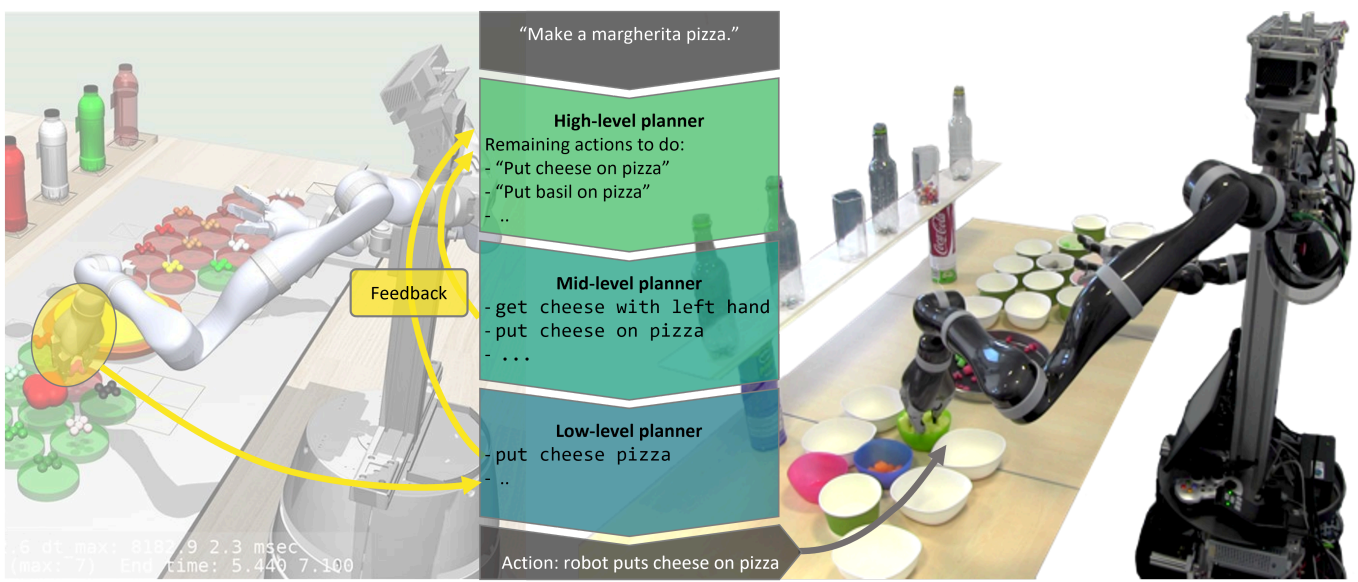

Snapshot of a pizza domain task execution on the virtual and real robot alongside an illustrative description of the CoPAL framework.

The implementation of CoPAL demonstrated significant improvements in the executability, correctness, and efficiency of robotic plans. The system's ability to replan and incorporate various types of feedback led to a higher success rate in completing tasks, reduced the number of planning errors, and improved the overall robustness and resilience of the robots in dynamic environments.

The empirical evaluations showed that CoPAL's feedback mechanisms and hierarchical planning structure effectively address the challenges of operating in complex, real-world scenarios.

The research and development of CoPAL with HyperGraphOS demonstrates the flexibility of modeling, executing, debugging, and testing complex data flow models that generate tasks for a humanoid robot in the real world. The core development of the multi-agent system, including the DSL, took only a single week, allowing most of the time to be spent on evaluation and experiments with the model.
In continuation of this work, HypergraphOS is now used for creating LLM-based multiagent system (MAS) architectures. A DSL for agents, multi-agent dialogue arbitration, working rooms and LLM accessible tools are currently the subject of implementation and research. This has been successfully demonstrated by flexible multi-party dialog generation and their analysis using a method also implemented on HypergraphOS **[Ebubechukwu, 2025]**.

# Case Study 3: Dynamic Research Projects with Thebes DSL

This case study addresses the challenge of dynamically evolving research projects, which often lack formalized requirements and undergo frequent changes. Traditional modeling languages and tools, such as SysML **[Hause, 2006]**, are too rigid and detailed, creating barriers to adoption and integration with other domain-specific tools.

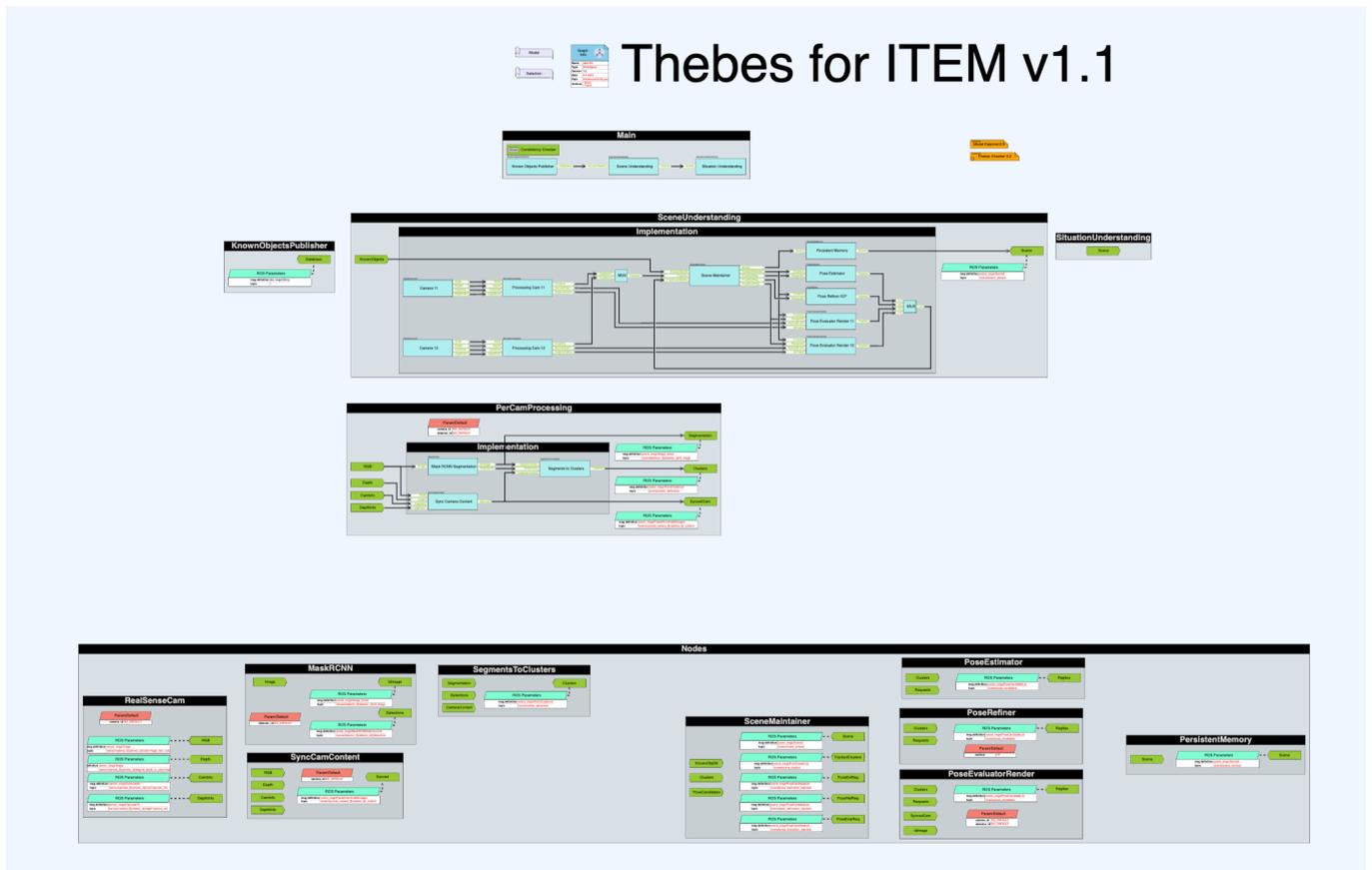

Model of a subset of the implementation of the HARU Project using Thebes Meta-Model

The work involves the development of a lightweight systems modeling language named Thebes, implemented as a HyperGraphOS DSL, specifically tailored for research projects. Thebes is designed to be easy to learn, frequently reconfigurable, and seamlessly integrable with existing tools through the use of code generators. This language supports incremental design and rapid prototyping by enabling researchers to manage system aspects without needing a full specification upfront.

The implementation of Thebes was demonstrated through two case studies: a robotics project for the table-top robot Haru and the CoPAL system, which uses natural language processing and large language models for interactive planning. The implementation of Thebes as a meta-model in HyperGraphOS showed significant benefits in managing and understanding system configurations and dynamics. For the Haru robotics project **[Gomez, 2018]**, the system model was constructed in one day, providing immediate insights and enabling better documentation and reasoning about the system. Similarly, in the CoPAL system **[Joublin, 2023]**, Thebes facilitated hierarchical composition and feedback loops, supporting dynamic system changes and interaction patterns. The case studies highlighted Thebes' effectiveness in reducing complexity, improving collaboration, and supporting rapid system modifications through code generation, ultimately enhancing the efficiency and productivity of research teams.

For the Haru robotics project, the system model was constructed in one day, providing immediate insights and enabling better documentation and reasoning about the system. Similarly, in the CoPAL system, Thebes facilitated hierarchical composition and feedback loops, supporting dynamic system changes and interaction patterns. The case studies highlighted Thebes' effectiveness in reducing complexity, improving collaboration, and supporting rapid system modifications through code generation, ultimately enhancing the efficiency and productivity of research teams.

## Case Study 4: The Making of This Article 😉

This case study has been included for completeness. We simply wish to take this opportunity to demonstrate the capabilities of HyperGraphOS by revealing that even the paper you are reading now was designed, drafted, edited, and rendered within HyperGraphOS. For this paper, we used a Multi-Agent Editor Office (MAEO) application that we previously developed in HyperGraphOS. The MAEO was conceived as a general-purpose AI-based system designed to assist writers in creating articles, books, and other written content.

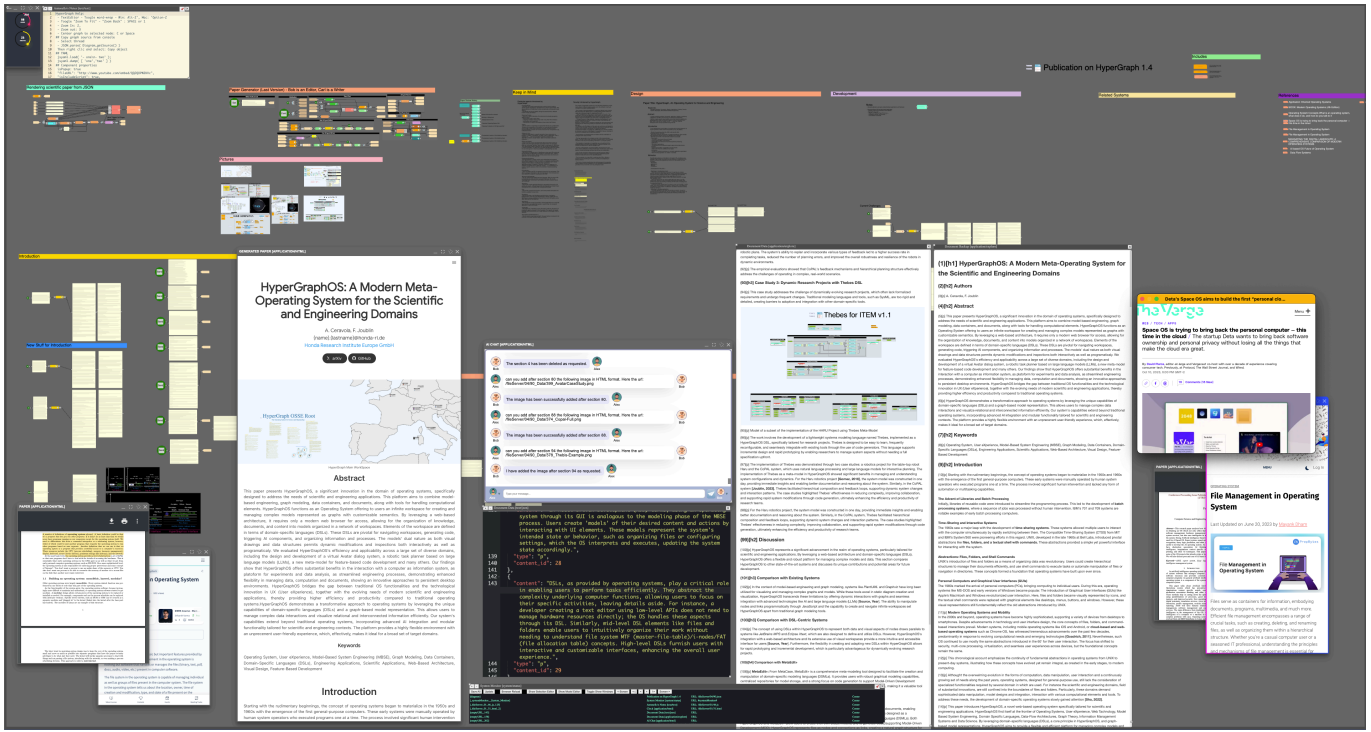

HyperGraphOS WorkSpace used for the full development of this article. The WorkSpace contains several models at the top (from left to right: paper renderer executable model, MAEO executable model — below which are some of the pictures from this article, design elements and drafts of the article, article versioning repository, notes, and related articles). On the left are several section drafts for the introduction, along with windows holding related articles, the paper render output, the chat with MAEO, and the raw content of the article (below you can see the system monitor), as well as the running and backup versions of the article and additional related articles.

This application has been used in several writing projects. The system supports the writer by organizing the structure of the document. The two main agents, 'Editor' and 'Writer,' have distinct roles. The 'Editor' focuses on the document's structure, managing insertions, deletions, and movement of paragraphs/sections, while the 'Writer' aids in formulating the content, offering suggestions and improvements for grammar and semantics. The creation of this particular multi-agent architecture, which provides text-editing tools to one agent, was developed in less than three days using the DataFlow DSL. The creation of the HyperGraphOS article is just one of many recursive experiences we've had with the system, including the redesign of the system itself.

# Discussion

HyperGraphOS represents a significant advancement in the field of operating systems, particularly for scientific and engineering applications. By leveraging a web-based architecture and domain-specific languages (DSLs), HyperGraphOS provides a flexible and robust platform for managing complex models and data. This section compares HyperGraphOS to other state-of-the-art systems, highlighting its unique contributions and identifying potential areas for future development.

**Comparison with Existing Systems**
In the context of model-based engineering and graph modeling, systems like *PlantUML* and *Graphviz* have long been used for visualizing and managing complex graphs and models. While these tools excel in creating static diagrams and visualizations, HyperGraphOS goes beyond these limitations by offering dynamic interactions with graphs and seamless integration with advanced technologies like AI and large language models (LLMs) **[Wei, 2022]**. The ability to programmatically manipulate nodes and links through JavaScript, along with the capacity to create and navigate infinite workspaces, distinguishes HyperGraphOS from traditional graph modeling tools.

### Comparison with DSL-Centric Systems

The concept of using DSLs within HyperGraphOS to represent both data and visual aspects of nodes is similar to systems like *JetBrains MPS* and *Eclipse Xtext*, which are also designed to define and utilize DSLs. However, HyperGraphOS's integration with a web-based architecture and its extensive use of visual workspaces provide a more intuitive and accessible interface for users. Additionally, the flexibility in creating and adapting new DSLs within HyperGraphOS facilitates rapid prototyping and incremental development, which is particularly beneficial for dynamically evolving research projects.

## Comparison with MetaEdit+

### Purpose and Scope

*MetaEdit+* **[Smolender, 1991][Tolvanen, 2006]**, developed by MetaCase, is a comprehensive meta-modeling tool designed to support the creation of domain-specific modeling languages (DSMLs) **[Kelly, 1996]** and code generation for Model-Driven Development (MDD) **[Tolvanen, 2003][kelly, 2008]**. It excels in specialized engineering environments, focusing on creating tailored modeling tools for specific domains like software engineering, systems engineering, and business process modeling. In contrast, HyperGraphOS goes beyond just modeling, combining DSLs with dynamic task execution, real-time collaboration, and AI integration. HyperGraphOS offers a more interdisciplinary approach, providing flexible, infinite workspaces that cater to a wider variety of tasks and domains, while *MetaEdit+* is specialized for detailed, structured modeling environments.

### Architecture

*MetaEdit+* relies on a centralized repository system, which ensures consistency and version control of models but may limit flexibility, especially in distributed or large-scale environments. It employs a proprietary language for defining models, which can introduce a steeper learning curve. On the other hand, HyperGraphOS adopts a distributed, web-based architecture that enables broad access to data and task management across local, cloud, and remote environments. The use of JavaScript as the primary shell language in HyperGraphOS provides programmatic control over nodes and links, offering an open, flexible development environment. This distributed architecture and reliance on a common language like JavaScript allow HyperGraphOS to integrate more seamlessly with external tools and workflows.

### Key Functionalities

*MetaEdit+* shines with its extensive graphical modeling tools, customized environments for different domains, and strong focus on MDD. It offers powerful modeling capabilities within a structured, repository-based system, making it ideal for large, complex engineering projects where model consistency and integrity are critical. However, HyperGraphOS moves beyond static models, offering dynamic task execution and real-time collaboration within its workspaces. While *MetaEdit+* provides strong support for model creation and code generation, HyperGraphOS enhances its utility by allowing users to work collaboratively, dynamically execute tasks, and integrate AI tools directly into the modeling process. HyperGraphOS also supports visual grouping and linking of documents in an infinite workspace, allowing for highly flexible, dynamic organization of data, in contrast to *MetaEdit+*'s more rigid model structure within repositories.

### Strengths and Limitations

*MetaEdit+* excels in providing a formalized, structured environment for model-driven development with a focus on rigorous validation and consistency checks. This makes it highly effective for users who need precise control over their modeling environment. However, the reliance on a centralized repository and proprietary language may limit scalability and flexibility, particularly in distributed environments.

HyperGraphOS, in contrast, emphasizes ease of use with its minimalistic, intuitive interface and dynamic capabilities. Its ability to integrate AI, offer real-time collaboration, and provide a flexible, distributed architecture gives it a distinct edge for interdisciplinary applications and evolving research projects. However, HyperGraphOS may not provide the same level of detailed, formal model validation and consistency as *MetaEdit+* yet.

## Comparison with ConceptBase

**Purpose and Scope**
*ConceptBase* **[Jeusfeld, 2018]** is a high-level conceptual modeling tool focused on meta-modeling, information modeling, and knowledge management **[Jeusfeld, 2006]**. It is designed to support complex models, particularly in fields like database design, software engineering, and organizational modeling. *ConceptBase* excels in environments where formalization and rigorous validation are paramount. On the other hand, HyperGraphOS emphasizes flexibility and dynamic interaction, providing a platform that integrates DSLs, AI tools, and real-time task execution within an infinite, collaborative workspace. While *ConceptBase* is suited for structured, logical modeling with a focus on consistency, HyperGraphOS caters to a broader range of tasks and applications, offering flexibility for interdisciplinary workflows.

**Architecture**
*ConceptBase* relies on a centralized repository architecture, ensuring model consistency and version control but potentially limiting flexibility in distributed environments. Its use of a Prolog-like language for querying and defining constraints ensures powerful, logic-based model manipulation but may require specialized knowledge. In contrast, HyperGraphOS adopts a distributed, web-based architecture, which allows users to store and manage WorkSpaces across various environments, such as local storage, cloud services, and remote computers. This distributed setup offers greater flexibility and easier data access. Moreover, HyperGraphOS uses JavaScript as its primary shell language, which offers programmatic control over models and the integration of external tools, making it more accessible to developers familiar with JavaScript and open systems.

**Key Functionalities**
*ConceptBase* provides strong support for defining and querying complex models, focusing on formal semantic validation and consistency checks. It is ideal for environments where rigor and accuracy are essential, offering an extensive querying system to analyze model interrelationships. HyperGraphOS, by contrast, is centered around dynamic interaction with models, offering advanced AI integration for real-time collaboration and task execution. HyperGraphOS supports the creation of infinitely flexible workspaces where documents and models can be grouped and linked, allowing for seamless navigation and modification. While *ConceptBase* is more focused on formal knowledge representation and querying, HyperGraphOS provides a more dynamic, user-driven experience with a broader focus on productivity and collaboration.

**Strengths and Limitations**
*ConceptBase* excels in supporting rigorous, formalized modeling with its powerful querying capabilities and logical framework. It ensures consistency and reliability, making it ideal for applications requiring structured, validated models, such as database design and organizational modeling. However, its centralized architecture and reliance on a specialized querying language can make it less flexible and harder to scale in distributed environments. HyperGraphOS, in contrast, offers dynamic, real-time task execution, infinite flexibility in data organization, and collaborative workspaces. Its distributed, web-based approach and ease of integration with external tools make it highly adaptable for evolving research and interdisciplinary applications. However, it may not provide the same level of formal model validation and consistency checks as *ConceptBase*, making it less suitable for applications requiring rigorous model validation.

# Comparison with Web-based Modelling Tools

## Comparison with WebGME

### Purpose and Scope
*WebGME* **[Maróti, 2014]** focuses on providing a web-based platform for creating and managing Domain-Specific Modeling Languages (DSMLs) with strong support for collaborative modeling. Its primary aim is to unify metamodeling and modeling tasks through a scalable infrastructure. In contrast, HyperGraphOS operates as a meta-operating system, combining graph-based structures with advanced Domain-Specific Languages (DSLs) and AI-driven tools. HyperGraphOS extends beyond modeling into dynamic task execution and interdisciplinary workflows, catering to a broader range of domains and applications.

### Architecture
*WebGME* is inherently web-based and optimized for scalability, featuring a backend that supports distributed version control and a frontend for visual modeling. It employs prototypical inheritance to allow efficient reuse and hierarchical organization of models. HyperGraphOS also utilizes a web-based architecture but emphasizes an infinite, interconnected workspace (OmniSpace) for seamless navigation across tasks and domains. This design allows it to integrate complex computations, including real-time AI assistance, into its workflows.

### Key Functionalities
*WebGME* offers functionalities such as version control for collaborative projects, prototypical inheritance to streamline model reuse, and support for multi-paradigm modeling across various domains. In comparison, HyperGraphOS provides dynamic task execution and real-time collaboration, integrates AI for modeling and execution support, and features advanced visualization and manipulation of graph-based structures.

### Strengths and Limitations
While *WebGME* excels in collaborative DSML development and version-controlled modeling, it lacks the dynamic task execution and AI-driven support that HyperGraphOS provides. HyperGraphOS's strength lies in its ability to adapt to interdisciplinary needs and support dynamic, real-time interactions.

## Comparison with jjodel

### Purpose and Scope
*jjodel* **[Di Rocco, 2014]** is a reflective, cloud-based modeling framework designed to simplify model-driven engineering by focusing on accessibility and real-time collaboration. Its primary audience includes educators and technical users in specific domains like railway engineering. HyperGraphOS's scope is broader, targeting complex interdisciplinary workflows and offering a highly extensible platform for creating and managing DSLs and executing tasks in real time.

### Architecture
*jjodel* leverages a reflective architecture to facilitate the co-evolution of metamodels and syntax definitions. Its cloud-based infrastructure emphasizes simplicity, "zero-setup" usability, and real-time collaborative features. HyperGraphOS shares the benefits of a cloud-based architecture but extends its functionality through a modular design that supports fractal extensibility. Its OmniSpace concept enables users to navigate infinite, interconnected workspaces, offering flexibility far beyond static modeling environments.

**Key Functionalities**

*jjodel* provides a reflective architecture for continuous co-evolution, intuitive syntax customization for domain-specific tasks, and a geometry-based notation for technical domains. In comparison, HyperGraphOS is also reflective but provides fractal extensibility to adapt and create new functionalities, integrates AI tools for advanced modeling and task automation, and supports multi-domain applicability for large-scale, adaptive systems.

**Strengths and Limitations**

*jjodel* excels in lightweight, collaborative modeling with a strong focus on accessibility and usability. However, its scope is narrower compared to HyperGraphOS, which supports interdisciplinary and computationally intensive workflows. HyperGraphOS's integration of AI and extensible design makes it a better fit for dynamic, large-scale applications.

## Comparison with Sirius Web

**Purpose and Scope**

*Sirius Web* **[Giraudet, 2024]** is a language workbench for creating graphical DSMLs and their environments. It builds on lessons learned from Sirius Desktop, focusing on collaborative modeling and usability. HyperGraphOS extends beyond the modeling scope of *Sirius Web* by acting as an operating system that supports both modeling and execution tasks, targeting applications that require dynamic, real-time capabilities.

**Architecture**

*Sirius Web* employs a web-native architecture with a backend developed in Java and Spring Boot and a frontend using TypeScript and React. It relies on GraphQL for communication between the server and client, ensuring seamless updates and collaboration. HyperGraphOS similarly adopts a web-based architecture but prioritizes seamless integration of distributed OmniSpaces for managing tasks and data. Its design allows infinite scalability and dynamic adaptability to user needs, combining real-time execution with intuitive modeling tools.

**Key Functionalities**

*Sirius Web* offers key functionalities such as predefined representation types (e.g., diagrams, Gantt charts), low-code and API-driven interfaces for creating modeling studios, and support for collaborative workflows. In comparison, HyperGraphOS provides dynamic modeling and task execution within a single platform, advanced AI integration for enhanced productivity, and infinite, flexible workspaces tailored to interdisciplinary applications.

**Strengths and Limitations**

*Sirius Web* excels in graphical DSML development and collaborative modeling but is primarily a tool for studio creation. HyperGraphOS, by contrast, operates as a more versatile platform that integrates modeling with task execution, making it suitable for complex, adaptive systems.

## Summary of Key Differentiators

| Feature | HyperGraphOS | WebGME | jjodel | Sirius Web | MetaEdit+ | ConceptBase |
|---|---|---|---|---|---|---|
| Primary Focus | New OS integrating Meta-modeling concepts | Collaborative DSML development tool | Lightweight meta-modeling tool | Graphical DSML tool | Meta-Modeling tool | Meta-modeling, information modeling, and knowledge management |
| Architecture | Web-Based, modular, reflective, fractal extensibility | Web-based, scalable version control | Cloud-based, reflective modeling | Web-based, GraphQL-based | Stand-alone application | Stand-alone application |
| Core Strengths | Novel OS with MBSE & AI integration | Multi-paradigm modeling, collaboration | Intuitive model customization | Predefined representations, collaboration | Visual DSL definition, mature product | Rigorous formalization of metamodel and model semantics |
| Target Users | Science and engineering | DSML developers and end-users | Educational and technical domains | DSML developers and end-users | DSL developers and end-users | DSML developers and end-users |

# Future Improvements

Although HypergraphOS already integrates AI at different level, these capabilities can be pushed forward towards complex task automation and code generation, enhancing productivity and efficiency. Comparatively, systems like *IBM Watson* and *Google AI* have showcased the immense potential of AI in transforming traditional workflows. HyperGraphOS could build on this potential by providing tailored AI interactions within each document, offering on-demand assistance and intelligent data manipulation.

While HyperGraphOS provides a high degree of flexibility and customization, its reliance on external services for security and privacy might pose challenges. Future iterations could explore the integration of more robust, built-in security measures to enhance user data protection. Additionally, expanding the scalability features to support even larger and more complex datasets without compromising performance will be crucial as the system evolves. Improving collaborative features is also on its agenda.

# Conclusion

In this paper, we have introduced HyperGraphOS, a modern web-based operating system designed specifically for scientific and engineering applications. Through leveraging domain-specific languages and a graph-based model representation, HyperGraphOS facilitates a more intuitive and flexible interaction with digital information, enabling users to create, manipulate, and visualize complex models and data **[Van Der Aalst, 2016]**.

Our case studies in virtual receptionist systems, robotic task planning, and dynamic research projects, multi-agent system or document manipulation and creation demonstrate the system's versatility and practical benefits. By comparing HyperGraphOS with other state-of-the-art systems, we have outlined its unique contributions and areas where it excels, such as dynamic interaction with graph models, flexibility in creating new DSLs, and seamless integration with AI components **[Warren, 2024]**.

While HyperGraphOS offers many advantages, it also opens up avenues for further enhancement, particularly in aspects of data handling, scalability, and integrated security measures. Future developments could focus on expanding these capabilities to accommodate more complex and larger datasets, as well as incorporating more robust security protocols.

Our work on HyperGraphOS addresses several challenges encountered over the years in research, system development, and prototyping. It aims to contribute to advancing the motivation for innovation in the field of Operating Systems. We have observed a stagnation in this domain, while simultaneously being perplexed by the substantial efforts directed towards the proliferation of computer languages and their rapid evolution. We firmly believe that instead of continuing to expand the variety of programming languages, we should focus on consolidating them to a minimal set, thereby creating more space for visual languages, which are well-known for bridging gaps between individuals with diverse expertise (for example, consider the widespread use of slides in presentations). Today, most computer languages have evolved towards a common computational model, making them nearly interchangeable. We propose that a more fruitful avenue lies in the development of Operating Systems that cater to the diverse needs of users. For instance, the recent proposal of Space OS by the startup Deta **[Pierce, 2023]** presents a new cloud-based operating system aimed at redefining personal computing. It gives users control over their applications and data in a personal cloud, offering a flexible and customizable environment where users can manage apps from any device. Additionally, the platform incorporates AI-powered tools like Teletype, enabling non-programmers to build applications through natural language. Another example is the Rabbit R1 **[Nield, 2024]**, a compact AI-powered device that simplifies user interaction with apps and services. It features a 2.88-inch touchscreen, a 360-degree rotating camera, and runs Rabbit OS, an AI-driven operating system. The R1 functions as a personal assistant, capable of performing tasks such as booking flights, playing music, or ordering food, all through voice commands. Powered by its Large Action Model (LAM), it provides faster responses and personalized functionality, positioning it as a next-generation alternative to traditional smartphone interfaces.

In summary, HyperGraphOS presents a novel approach to operating system design that aligns with the evolving needs of modern scientific and engineering workflows, providing a flexible, efficient, and user-friendly platform for managing intricate models and data.